Up- and Down-Conversion between Intra- and Inter-Valley Excitons in Waveguide Coupled Monolayer $WSe_2$

Yueh-Chun Wu,[1] Sarath Samudrala,[2] Andrew McClung,[2] Takashi Taniguchi,[3] Kenji Watanabe,[4] Amir Arbabi[2,*] and Jun Yan[1,*]

[1]Department of Physics, University of Massachusetts Amherst, Amherst, Massachusetts 01003, USA

[2] Department of Electrical and Computer Engineering, University of Massachusetts Amherst, Amherst, Massachusetts, 01003, USA

[3] International Center for Materials Nanoarchitectonics, National Institute for Materials Science, Tsukuba, Ibaraki 305-0044, Japan

[4] Research Center for Functional Materials, National Institute for Materials Science, Tsukuba, Ibaraki 305-0044, Japan

[*]Corresponding Authors: Amir Arbabi, arbabi@ecs.umass.edu; Jun Yan, yan@physics.umass.edu.

**ABSTRACT:**

The presence of two spin-split valleys in monolayer (1L) transition metal dichalcogenide (TMD) semiconductors supports versatile exciton species classified by their spin and valley quantum numbers. While the spin-0 intravalley exciton, known as the 'bright' exciton, is readily observable, other types of excitons, such as the spin-1 intravalley (spin-dark) and spin-0 intervalley (momentum-dark) excitons, are more difficult to access. Here we develop a waveguide coupled 1L tungsten diselenide ($WSe_2$) device to probe these exciton species. In particular, TM coupling to the atomic layer's out-of-plane dipole moments enabled us to not only efficiently collect, but also resonantly populate the spin-1 dark excitons, promising for developing devices with long valley lifetimes. Our work reveals several upconversion processes that bring out an intricate coupling network linking spin-0 and spin-1 intra- and inter-valley excitons, demonstrating that intervalley scattering and spin-flip are very common processes in the atomic layer. These experimental results deepen our understanding of tungsten diselenide exciton physics and illustrate that planar photonic devices are capable of harnessing versatile exciton species in TMD semiconductors.

Monolayer transition metal dichalcogenide (1L-TMD) semiconductors are a class of two-dimensional (2D) materials with exceptional optical properties. The monolayer, only three atoms thick, can have as much as 15% optical absorption.[1,2] The direct gap system has a very fast radiative recombination rate[3,4] that enables strong photoluminescence (PL) emission, thanks to the tightly bound nature of its excitons.[5–8] At the bandgap, 1L-TMDs possess two distinct valleys that are spin-split (Figure 1a), as a result of its broken inversion symmetry.[9,10] These valleys are fortuitously located at the high-symmetry corner points of the hexagonal Brillouin zone, enabling convenient optical access to the spin and valley degrees of freedom[11] for manipulation of the valley polarization and coherence.[12]

Considering an individual valley, aside from the spin-0 intravalley exciton X (Figure 1a), widely known as the 'bright' exciton due to its prominent coupling to optical fields, 1L-TMD excitons also possess a spin-1 dark sector $X_d$ (Figure 1b). This 'spin-dark' intravalley exciton, with zero in-plane electric dipole moment, is a much weaker optical feature than X. However, it has important implications for the optical properties of tungsten-based TMDs such as 1L-WSe$_2$. As can be seen from Figures 1a and 1b, in 1L-WSe$_2$ the dark exciton has a lower energy than the bright exciton, since the lowest conduction band and the highest valence band have opposite spins.[9] The bright-dark splitting is ~40 meV;[13–18] thus at low temperatures (LT), the dark exciton is much more populated than the higher-energy bright exciton. This leads to quenching of the bright exciton PL intensity,[19] in contrast to, *e.g.*, MoSe$_2$, whose PL emission becomes more intense at lower temperatures.[20,21] The prevalence of dark excitons at LT makes them readily available to bind to a bright exciton or a trion, forming large complexes containing four and five particles with intriguing spin and valley configurations.[22–26] Mediated by the spin-orbit interaction, $X_d$ does acquire an out-of-plane electric dipole moment,[15,27] although the strength is much weaker than that for the in-plane dipole of X. This weak strength, together with its role as the two-particle ground state in 1L-WSe$_2$, result in $X_d$'s much longer lifetime.[28] Further, due to its spin configuration (Figure 1b), the dark exciton is not prone to valley depolarization through the Coulomb exchange interaction,[29] making it an interesting candidate for valleytronics.[16,17] So far population of dark excitons relied on off-resonance excitation with photons at higher energy, which unavoidably incurs significant losses such as radiative recombination as the bright exciton. Presumably, it would be more efficient to populate the dark exciton directly by pumping at its resonance energy. This is challenging however, since the dark exciton does not absorb photons with electric field polarized in the plane, a prevailing experimental condition in backscattering geometry.

Another type of low energy exciton of relevance is the spin-0 intervalley exciton $X_i$ with the electron and hole residing in two opposite valleys (Figure 1c). This exciton state is even more elusive than $X_d$ and it evades typical optical spectroscopy studies due to its large momentum mismatch with photons. $X_i$ is expected to have important impacts on the optical properties of 1L-WSe$_2$, since its energy is also smaller than that of the bright exciton.[19] From time-reversal symmetry, $X_i$ has the same band energy as $X_d$, but this degeneracy is lifted by exchange interaction, which raises the intervalley exciton energy slightly. Recent experiments found that $X_i$ is about 10 meV above $X_d$.[30,31] Despite this increase, the energy difference between X and $X_i$ is still larger than a typical K point phonon.[32–34] It is thus reasonable to speculate that $X_i$ forms an important decay channel for the bright exciton through the emission of K point phonons.

In this work we use photoluminescence excitation (PLE) spectroscopy, a sensitive technique that has been previously employed to reveal weak features such as the excited Rydberg excitons in TMDs,[35–37] to probe these more elusive exciton species. To be able to access both the in-plane (IP) and the out-of-plane (OP) dipole moments of the atomic layer, we couple 1L-WSe$_2$ to a silicon nitride (SiN$_x$) waveguide that is designed to support well-defined TE and TM modes over a broad spectral range. In particular, TM coupling into the OP dipoles enabled us to resonantly populate dark excitons in the atomic layer with proper polarization. Under such experimental conditions, up- and down-conversion processes from the dark exciton to other exciton species become available to our study. Enhanced dark exciton detection efficiency taking advantage of the TM mode propagation in the waveguide also enabled us to investigate processes that can absorb energy in the atomic layer to generate up-converted dark excitons. These measurements bring out an interaction network between various TMD excitons and showcase a simple planar integrated optical microchip that exploits versatile TMD exciton modes with different quantum numbers.

**RESULTS AND DISCUSSION**

The 1L-WSe$_2$ is exfoliated from bulk crystals grown using chemical vapor transport method and sandwiched between hexagonal boron nitride (hBN) with a dry transfer technique, similar to our previous studies.[8,22,37] The hBN encapsulated device is transferred to silicon nitride ($SiN_x$) waveguide fabricated on a 2.5-µm-thick thermally grown $SiO_2$ on silicon substrate (Figure 1d). The waveguide core is 1 µm by 250 nm and operates in the wavelength range of 400 nm to 1550 nm. The waveguide was fabricated by depositing a 250-nm-thick layer of $SiN_x$ on the oxidized silicon wafer, followed by electron beam lithography patterning and dry etching. The electric field distributions for the TE and TM modes of the waveguide at 730 nm (1.7 eV) are shown in Figure 1f. For TE/TM, the fields are dominantly in the *x*/*y* (or H/V) directions. The waveguide coupled device is loaded in a cryostat with the exposed waveguide facet facing the optical window. The distinctive waveguide mode patterns enable us to selectively couple to TE and TM modes with H and V polarization configurations. We use an objective lens with NA=0.35 to couple incident radiation into the waveguide and collect PL emission from 1L-$WSe_2$.

Figure 2a shows polarization-resolved PL spectra of our device at 3.4 K under 532 nm CW laser illumination. With H polarized excitation, we selectively collect the PL signal polarized at an angle $\theta$ with respect to the incident polarization ($\theta = 0°/90°$ corresponds to H/V collection respectively). Seven well-resolved peaks are observed, with X the bright neutral exciton (Figure 1a), BE the biexciton,[22] $T_1$ and $T_2$ intervalley and intravalley trions, $X_d$ the dark exciton (Figure 1b), ET the five-particle exciton-trion,[22] $X_i^r$ a phonon replica [30,31,38] of the intervalley exciton $X_i$ (Figure 1c). These peaks are similar to our previous devices studied in backscattering geometry,[22] demonstrating that the coupling with waveguide did not cause much degradation to sample quality. The simultaneous appearance of neutral (X, BE, $X_d$, $X_i^r$) and charged species ($T_1$, $T_2$, ET) indicates that the device is slightly electron-doped.[8,22,37] We note that in this low electron density doping regime, $X_i^r$ is accidentally dengenrate[38] with the negatively-charged dark trion $T_d$.[13,39] Thus the peak at ~1.67eV likely has some contribution from $T_d$ especially in the V-polarized collection. The $\theta$ dependent intensity profile (Figure 2b) clearly shows that the dark exciton's selection rule is opposite to the other six emission features: the dark exciton is more sensitive to the V polarization while the others are more sensitive to H. In fact, in V collection $X_d$ becomes the most intensive emission feature, consistent with TM selection of the OP dipole moment. These off-resonance measurements are further substantiated in the PLE mapping where we resonantly excite the bright X exciton with H polarization. As shown in Figures 2c and 2d, in H collection no visible $X_d$ emission is observed while for V collection the $X_d$ emission dominates in the map. The resonance profile is seen to peak exactly at the bright exciton X energy 1.733 eV. This suggests that the spin-0 bright exciton can flip its electron spin and efficiently relax to the spin-1 dark exciton channel.

We subsequently use our device to explore the opposite process: can the dark excitons be resonantly populated and upconvert to the bright exciton? Exciton up-conversions were observed in TMDs before;[40–43] however, those processes do not involve interactions that drive a spin-flip. Our study here is, in fact, an interesting test of the spin-flip process alluded above. As shown in Figures 3a & 3b, up-conversion to the bright exciton is observable, but only in the VH, and not in the HH configuration. This shows that the dark excitons only absorb photons of TM polarization. In Figure 3c, we have plotted the VH resonance profile (red squares) together with a PL spectrum in the background. The upconverted bright exciton intensity is seen to peak exactly at the energy of $X_d$, indicating negligible Stokes shift between absorption and emission in our device. The observed resonant dark exciton absorption necessarily involves flipping the electron spin, assisted by the spin-orbit interaction, in accordance with theoretical calculations.[27] This resonant up-conversion from $X_d$ to X echoes the down-conversion map in Figure 2d, showing that there is a direct link between the population of dark and bright excitons, providing concerted support for the spin-flip process that balances the population between the two exciton species.

Given that the dark exciton couples only to OP polarized optical fields as demonstrated in Figures 2c & 2d for emission, and Figures 3a & 3b for absorption, we use in the following only V polarization when measuring at the $X_d$ energy. Now let's explore the types of excitonic modes that can be generated through down-conversion from the dark excitons. In Figures 3d and 3e, we use V polarized laser light to excite dark excitons in the atomic layer and collect the PL emission at lower energies. Interestingly we observed emission at ~1.67 eV in both H and V polarizations. We attribute the H emission to the chiral phonon replica $X_i^r$ [30,31,38] of the intervalley exciton as observed in Figure 2a. The V polarized emission at about the same energy likely comes from the dark trion $T_d$, considering its accidental degeneracy in energy with $X_i^r$,[38] as noted in the discussion of Figure 2a. It could also have a minor contribution from $X_i^r$ emission due to the limited extinction ratio of our waveguide coupling in selecting OP signals. As seen in Figure 1f, TM mode still has some residual in-plane electric field distribution along the length of the waveguide, making coupling to IP signal possible.

To gain a better understanding of the excitonic features at around 1.67 eV, we sweep the laser excitation in this energy range and performed up-conversion measurements to collect the V polarized dark exciton emission. From PLE maps in Figures 4a and 4b, both H and V excitations are capable of generating upconverted dark excitons. This is in contrast to results in Figures 3a and 3b, where only V absorption of the dark exciton is observed. The HV PLE is attributable to the upconversion from $X_i^r$ to $X_d$. We note that this is a somewhat nontrivial observation in light of previous studies. In the upconversion from the $T_1$ trion mode to bright exciton,[40] pseudoangular momentum (PAM) preserving out-of-plane chalcogen (OC) vibration[34] of $A_1'$ symmetry was believed to assist the transition. However, it was explicitly demonstrated that no X-$A_1'$ phonon replica was contributing to the upconverted signal. Here, in contrast, we find that the exciton-phonon replica $X_i^r$ can indeed absorb photons and create upconverted excitons. We further note that this upconversion is another process that requires a spin-flip for the electron involved since $X_i$ has the same spin as X.

For the VV map in Figure 4b, an interesting observation is that the upconverted $X_d$ intensity is more intense than that in the HV map. As seen in Figure 4c, VV intensity is about three times that of HV. This result rules out the possibility that VV upconversion is due to $X_i^r$ generation with non-ideal mode coupling as $X_i^r$ absorbs dominantly H polarized radiation; instead, this is a dark trion to dark exciton upconversion. It also corroborates that the VV down-conversion in Figure 3e is due mainly to dark exciton to dark trion transition. The dark trion to dark exciton upconversion in Figure 4b is a process similar to the bright trion to bright exciton transition observed in Ref. [40]. Here light absorption and spin-flip act together with an electron in the atomic layer to resonantly generate dark trions. The dark trions subsequently absorb energy in the system to disassociate the additional charge and produce neutral dark excitons, which make use of spin-orbit interaction to flip its electron spin and radiatively recombine. Thus this is a process that involves active participation of spin flips.

The multiple PLE up- and down-conversion processes observed in Figures 2-4 provide a comprehensive interrogation on the interplay between the three exciton species illustrated in Figure 1. We have summarized these relationships in Figure 5: the bright excitons are capable of generating intravalley dark excitons by spin-flip, and intervalley excitons by the emission of a K-point phonon. The dark exciton can flip its electron spin, harness energy in the system and upconvert to the bright exciton. Our observation of the $X_d$-$X_i^r$ down-conversion and the $X_i^r$-$X_d$ up-conversion further suggests that the dark exciton and the intervalley exciton can convert to each other efficiently. Thus, our experimental results show that, despite that the three exciton species in Figure 5 have different spin and/or valley characters, TMD atomic layers are quite efficient in switching these quantum numbers.  Even at cryogenic temperatures where large momentum phonon population is low, excitons are capable of harnessing tens of meV to upconvert to another type of exciton, and switch its valley and/or electron spin if needed. Lastly, we take note that not observed in the current work are the intervalley spin-flip exciton and the indirect exciton composed of a hole at K point and an electron at the halfway between K and Brillouin zone center. These excitons also have energies lower than the bright exciton[44] and most likely coexist with the three types of excitons shown in Figure 5.

**CONCLUSIONS**

In conclusion, efficient coupling to IP and OP dipole moments of 1L-$WSe_2$ using a silicon nitride waveguide provides a powerful tool to investigate multiple TMD exciton species. We demonstrate that the resonant population of the spin-dark intravalley exciton is possible for devising devices with long valley lifetime. Our work revealed three up-conversions not visible before: from dark exciton to bright exciton, from dark trion to dark exciton, and from intervalley exciton phonon replica to dark exciton, substantiating that TMDs are highly efficient in harnessing energy from the environment even at low temperatures. We also show that spin-flip is a critical and frequently-occurring process in TMD atomic layers that participate in the resonant optical generation and recombination of dark excitons, dark trions, and in the interconversion between the spin-zero X, $X_i$ and spin-one $X_d$. The comprehensive PLE up- and down- conversion studies show that versatile exciton species coexist in 1L-$WSe_2$, and they are quite effective in harnessing energies, spin-flip interactions, and electron-phonon couplings in the system to transition into and balance with each other.

**METHODS**

**Crystal growth.** The bulk $WSe_2$ crystals are grown by the chemical vapor transport (CVT) method. High purity W 99.99%, Se 99.999%, and $I_2$ 99.99% (Sigma Aldrich) are placed in a fused silica tubing that is 300 mm long with an internal diameter of 18 mm. W and Se are kept in a 1:2 stoichiometric ratio with a total mass of 2 g. Sufficient $I_2$ is added to achieve a density of 10 mg cm$^{-3}$. The tube is pump-purged with argon gas (99.999%) for at least five times and sealed at low pressure prior to growth. Using a three-zone furnace, the reaction and growth zones are set to 1055 and 955 ℃, respectively. The growth time is approximately 2 weeks.

**Sample fabrication.** The atomic flakes of $WSe_2$, hexagonal boron nitride (hBN) are first exfoliated on Si wafers with 300 nm of $SiO_2$ and inspected under an optical microscope. To make the high-quality 1L-$WSe_2$ heterostructures, we annealed hBN in $O_2$/Ar atmosphere (5sccm/50sccm) at 500 ℃ for 2-3 hours before stacking. The flakes are then stacked and transferred on substrates with premade $SiN_x$ waveguides using a dry transfer technique with PPC (poly-propylene carbonate) stamp. All the exfoliation, inspection, and stacking processes are completed in a nitrogen-purged glovebox to minimize sample degradation. The waveguide sample is thermally annealed at 350 ℃ for 1 hour in argon environment to improve the quality.

**Photoluminescence excitation spectroscopy.** The sample is transferred to a closed-loop cryostat with optical access. The waveguide sample is mounted using GE-varnish with the exposed waveguide facet facing the optical window. All measurements in this work are performed at 3.4 K. A 532 nm (2.33 eV) laser is used for off-resonance excitation. For resonant excitations, we use a tunable Ti-sapphire laser together with a 4f two-grating spectral filter system. The incident laser is reflected by a nonpolarizing cube beam splitter and then focused on the sample by a 50× objective lens (NA: 0.35) with a spot size of ~2 μm. The incident beams are linearly polarized for coupling to TE and TM waveguide modes. The bright exciton density generated in our device is ~$10^8 – 10^9$ cm$^{-2}$, and the dark exciton ~$10^{10} – 10^{11}$ cm$^{-2}$. The collected optical signal is then guided into a Horiba T64000 spectrometer equipped with a liquid-nitrogen-cooled CCD camera.

**ACKNOWLEDGMENTS**

This work is supported mainly by the University of Massachusetts Amherst, and in part by NSF ECCS-1509599. K.W. and T.T. acknowledge support from the Elemental Strategy Initiative conducted by the MEXT, Japan, Grant Number JPMXP0112101001, JSPSKAKENHI Grant Numbers JP20H00354 and the CREST(JPMJCR15F3), JST.

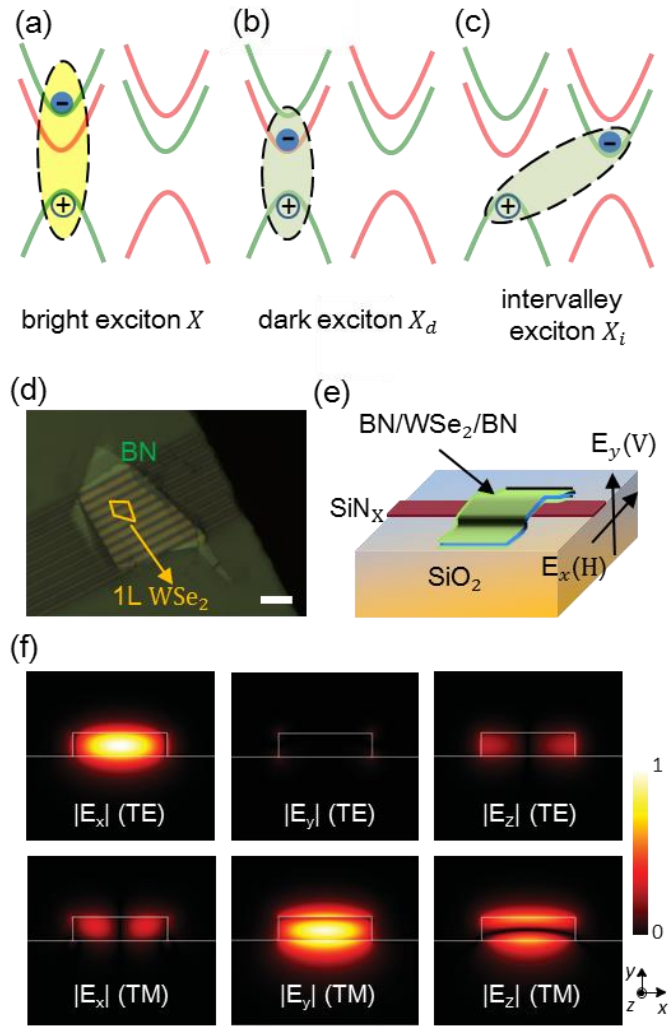

Figure 1. (a) Intravalley bright exciton $X$, (b) Intravalley dark exciton $X_d$, and (c) Intervalley exciton $X_i$ in 1L-WSe$_2$. (d) Sample picture (scale bar: 10 µm) and (e) schematic drawing of SiNx waveguide coupled hBN/1L-WSe$_2$/hBN device. The bright exciton $X$ is sensitive to H polarized optical field, while the dark exciton $X_d$ is sensitive to V. (f) Electric field distribution of the SiN$_x$ waveguide TE and TM modes at 730 nm. Waveguide dimension: 1 μm × 250 nm.

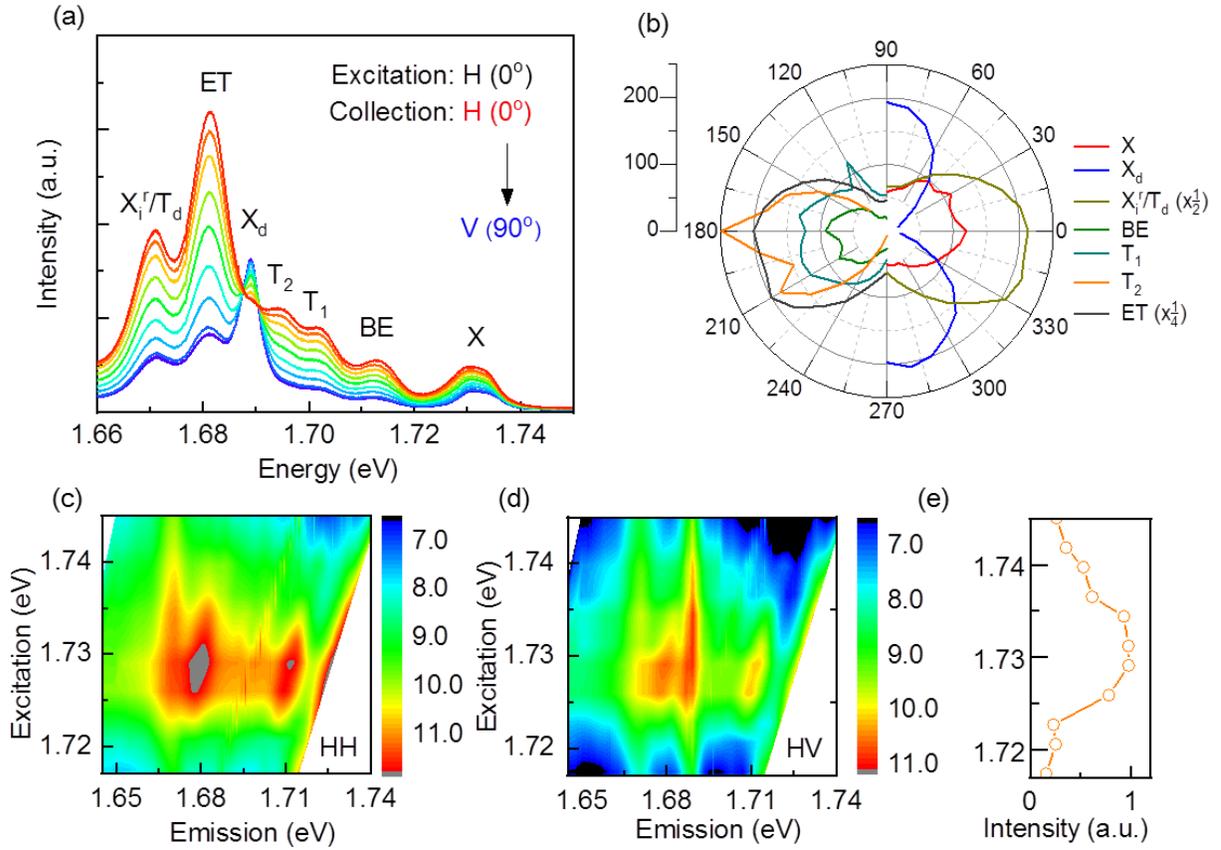

Figure 2. Polarization-resolved PL spectra of hBN/1L-WSe$_2$/hBN sample on waveguide at 3.4 K. (a) PL spectra with 532 nm H excitation at different collection angles from 0º (H) to 90º (V) in steps of 10º. Laser power: 150 µW. (b) The angular dependence of integrated intensity for the seven excitonic peaks observed in (a). (c) PLE contour plots in H-excitation/H-collection and (d) H-excitation/V-collection channels near bright exciton resonance. Intensity of contour plots is shown in natural logarithm scale. Laser power: 260 µW. (e) Normalized dark exciton intensity profile (linear scale) in (d).

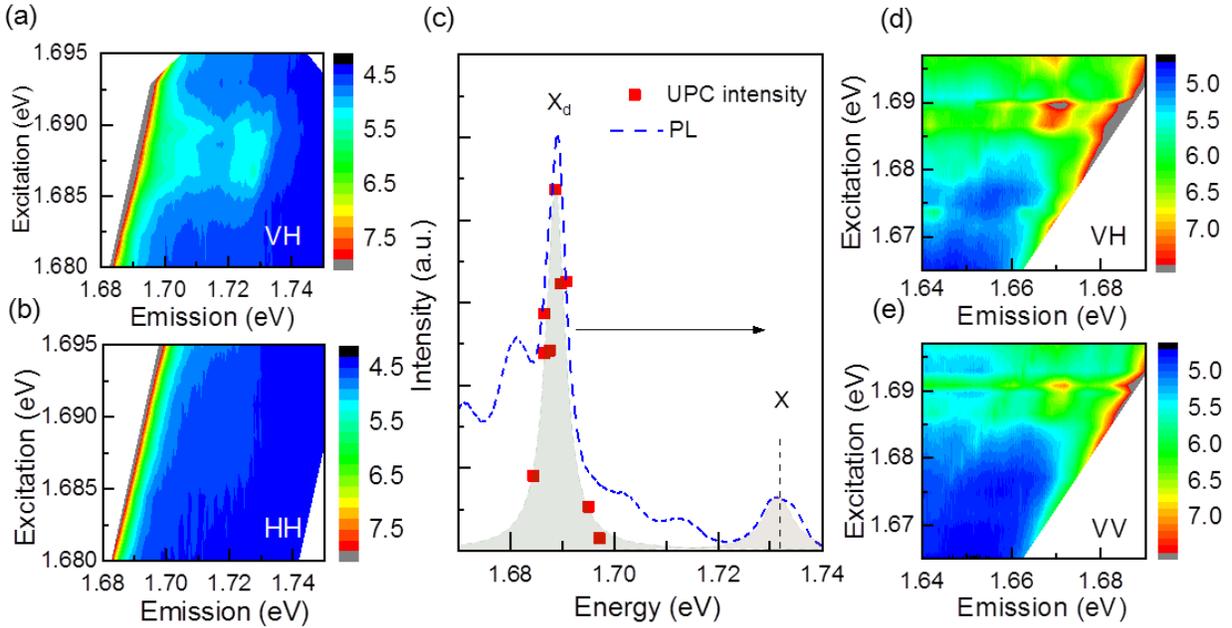

Figure 3. PLE scanning over the dark exciton resonance. (a) Anti-Stokes PLE contour plots in VH and (b) HH configurations. The up-conversion peak can be seen only with V excitation. (c) Integrated up-conversion emission intensity profile *vs.* excitation energy. The dashed blue curve shows a PL spectrum by 532 nm excitation where the bright (X) and dark ($X_d$) exciton emission peaks can be resolved. (d) Stokes PLE contour plots in VH and (e) VV configurations around dark exciton resonance. All spectra were taken with 550 µW laser power.

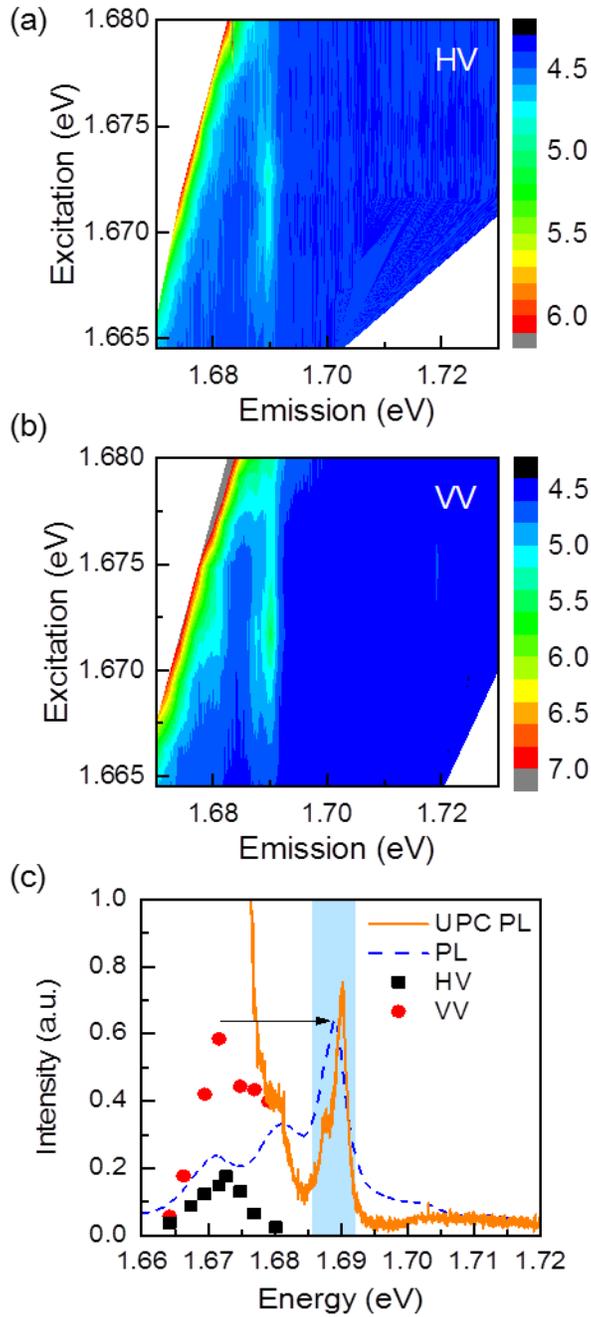

Figure 4. PLE around 1.67 eV excitation. (a) PLE upconversion contour plots in HV and (b) VV configurations. (c) Up-converted (UPC) PL intensity *vs.* excitation photon energy for HV (black square) and VV (red circle) polarizations. UPC PL due to 1.671 eV (orange solid line) excitation and regular PL spectrum by 532 nm excitation (blue dash line) are also plotted. The shaded region indicates the dark exciton emission. All spectra were taken with 550 µW laser power.

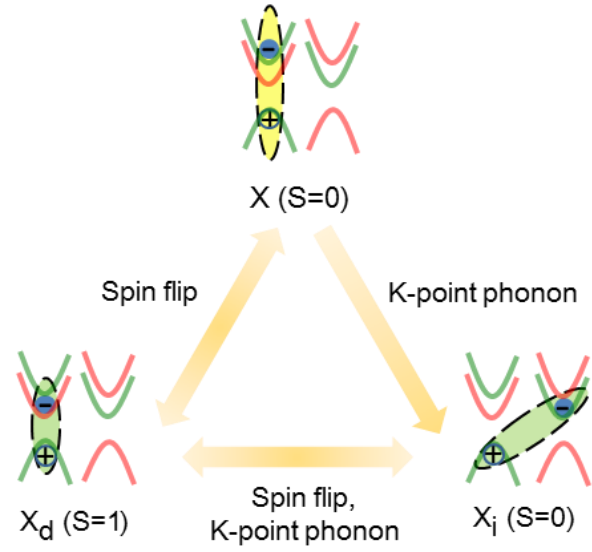

Figure 5. Schematic of up-conversion and down-conversion between the bright exciton, the dark exciton, and the intervalley exciton in monolayer WSe$_2$.

TOC figure

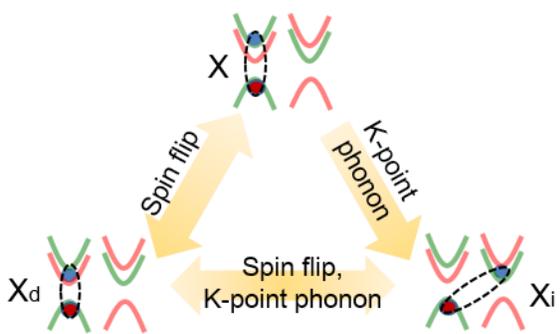